Comment on; arXiv:2510.13767; Structural origin of resonant diffraction in $RuO_2$

Stephen W Lovesey

Occhialini *et al*. [1] have added x-ray Bragg diffraction patterns on a single crystal of $RuO_2$ to an existing body of recent studies of its magnetic properties [2-5]. A consensus view of a non-magnetic symmetry has acquired significant substance. Specifically, the magnetic symmetry of the ruthenate is no longer thought to be the same as $MnF_2$, namely, $P4_2'/mnm'$ (No. 136.499 [6]). This symmetry prohibits the interference of non-magnetic diffraction (Templeton & Templeton (T&T) [7]; renamed in [1] as "structural") and magnetic diffraction in resonant x-ray Bragg spots. For, the corresponding amplitudes are separated by a phase of 90º, and T&T and magnetic x-ray intensities are in quadrature. As background information, $MnF_2$ nuclear and magnetic neutron scattering amplitudes are of one phase [8, 9], and likewise T&T and magnetic amplitudes for resonant x-ray diffraction by a linear magnetoelectric material [10]. Returning to Occhialini *et al*. [1], intensities cited in the main text and SM (Eqs. 14 & 16) violate magnetic symmetry $P4_2'/mnm'$, which is contrary to several statements by the authors [11]. Footnote SM [54] is a falsehood, because the mentioned interference of structural origin (T&T) is identically zero according to the assertion of $MnF_2$-type symmetry in data analysis. In which case, a correct conclusion is that observed intensity at the reflection vector (0, 0, 1) is consistent with pure T&T diffraction, with no evidence of axial magnetic dipoles permitted in $MnF_2$-type symmetry $P4_2'/mnm'$. Previously published symmetry informed x-ray Bragg diffraction amplitudes for $P4_2'/mnm'$ demonstrate that intensities are in quadrature for the specific reflection vectors (1, 0, 0) and (0, 0, 1) used by Occhialini *et al*. [12]. Moreover, it is a chiral magnetic symmetry, and Bragg spots intensities change on reversing the helicity of primary x-rays.

[1] C. A. Occhialini, C. Nelson, A. Bombardi, S. Fan, R. Acevedo-Esteves, R. Comin, D. N. Basov, M. Musashi, M. Kawasaki, M. Uchida, H. You, J. Mitchell, V. Bisogni, C. Mazzoli and J. Pelliciari. Structural origin of resonant diffraction in $RuO_2$. arXiv:2510.13767. Phys. Rev. B **113**, L060411 (2026).

[2] G. Yumnam, P. R. Raghuvanshi, J. D. Budai, D. Bansal, L. Bocklage, D. Abernathy, Y. Cheng, A. Said, I. I. Mazin, H. Zhou, B. A. Frandsen, D. S. Parker, L. R.Lindsay, V. R. Cooper, M. E. Manley and R. P. Hermann. Constraints on magnetism and correlations in $RuO_2$ from lattice dynamics and Mössbauer spectroscopy. https://doi.org/10.21203/rs.3.rs-6572669/v1 (2025).

[3] M. Hiraishi, H. Okabe, A. Koda, R. Kadono, T. Muroi, D. Hirai and Z. Hiroi. Nonmagnetic ground state in $RuO_2$ revealed by muon spin rotation. Phys. Rev. Lett. **132**, 166702 (2024).

[4] L. Kiefer, F. Wirth, A. Bertin, P. Becker, L. Bohat, K. Schmalz, A. Stunault, J. A. Rodríguez-Velamazan, O. Fabelo and M. Braden. Crystal structure and absence of magnetic order in single-crystalline $RuO_2$. J. Phys.: Condens Matter **37**, 135801 (2025).

[5] P. Keßler, L. Garcia-Gassull, A. Suter, T. Prokscha, Z. Salman, D. Khalyavin, P. Manuel, F. Orlandi, I. I. Mazin, R. Valentí and S. Moser. Absence of magnetic order in $RuO_2$: insights from µSR spectroscopy and neutron diffraction. NPJ Spintronics **2**, 50. https://doi.org/10.1038/ s44306-024-00055-y (2024).

[6] Belov-Neronova-Smirnova (BNS) setting of magnetic space groups, Bilbao Crystallographic server, http://www.cryst.ehu.es.

[7] D. H. Templeton and L. K. Templeton. Tensor X-ray optical properties of the bromate ion. Acta Cryst. A **41**, 133-4 (1985).

[8] H. A. Alperin, P.J. Brown, R. Nathans and S. J. Pickart. Polarized neutron study of antiferromagnetic domains in $MnF_2$. Phys. Rev. Lett. **8**, 237 (1962).

[9] Z. Yamani, Z. Tun and D. H. Ryan. Neutron scattering study of the classical antiferromagnet $MnF_2$. Canadian J. Phys. **88**, 771-797 (2010).

[10] S. W. Lovesey, E. Balcar, K. S. Knight and J. Fernández-Rodríguez. Electronic Properties of Crystalline Materials Observed in X-Ray Diffraction. Phys. Rep. **411**, 233 (2005).

[11] Z. H. Zhu, J. Strempfer, R. R. Rao, C. A. Occhialini, J. Pelliciari, Y. Choi, T. Kawaguchi, H. You, J. F. Mitchell, Y. Shao-Horn and R. Comin. Phys. Rev. Lett. **122**, 017202 (2019).

[12] S. W. Lovesey, D. D. Khalyavin and G. van der Laan, Phys. Rev. B **105**, 014403 (2022).